\pdfoutput=1
\documentclass[cernpreprint, atlasdraft=false, UKenglish, orcidlogo, texmf, orcidlogo]{atlasdoc}
 
\usepackage[subfig]{atlaspackage}
\usepackage{atlasbiblatex}
 
\usepackage{atlasphysics}

\graphicspath{{logos/}{figures/}}
 
\usepackage{ANA-BPHY-2020-07-PAPER-defs}
 
\usepackage{multirow}
\usepackage{proofread}
\usepackage{xcolor}
 
\usepackage{caption}
 
\usepackage{adjustbox}
\usepackage{braket}
\usepackage{float}

\AtlasTitle{Measurement of the $B_s^0 \to \mu\mu$ Effective Lifetime with the ATLAS Detector}

\AtlasAbstract{
This paper reports the first ATLAS measurement of the $B_s^0 \to \mu\mu$ effective lifetime. The measurement is based on the data collected in 2015-2016, amounting
to 26.3 fb$^{-1}$ of 13 TeV LHC proton-proton collisions.
The proper decay-time distribution of $58\pm13$ background-subtracted signal candidates is fit with simulated signal templates parameterised
as a function of the $B_s^0$ effective lifetime, with statistical uncertainties extracted through a Neyman construction. The resulting effective measurement of the
$B_s^0 \to \mu\mu$ lifetime is $0.99^{+0.42}_{-0.07} \, (\text{stat.})\pm 0.17 \text{ (syst.)}\,\mathrm{ps}$ and it is found to be consistent with the Standard Model.
}
 
\author{The ATLAS Collaboration}
 
\AtlasRefCode{BPHY-2020-07}
 
\PreprintIdNumber{CERN-EP-2023-139}

\AtlasJournal{JHEP}
\AtlasJournalRef{JHEP 09 (2023) 199}
\AtlasDOI{10.1007/JHEP09(2023)199}

\hypersetup{pdftitle={ATLAS document},pdfauthor={The ATLAS Collaboration}}
 
\begin{document}

\maketitle
 
\tableofcontents

\section{Introduction}
The Standard Model (SM) predicts that only the $CP$-odd heavy-mass eigenstate in the $B^0_s -\bar{B}^0_s$ pair decays into a di-muon final state \cite{Beneke_Power_enhance_SM_Bmumu_err, new_phys_via_eff_lifetime}. This exclusivity does not generally hold when considering
contributions Beyond the Standard Model (BSM) such as minimal supersymmetric Standard Model extensions \cite{np_in_flavour}, that can potentially perturb the effective
lifetime in \bsmumu\ decays. These perturbations can be significant also in absence of measurable BSM effects on the \bsmumu\ branching fraction (BR). The effective \bsmumu lifetime is defined as
$\taumumu=\frac{\int_0^\infty t \Gamma\left(B_s^0\left(t\right)\to\mu\mu\right)\, dt}{\int_0^\infty \Gamma\left(B_s^0\left(t\right)\to\mu\mu\right)\, dt}$, where $t$ is the proper decay time of the $B_s^0$ and $\bar{B}_s^0$ mesons and
$\Gamma\left(B_s\left(t\right)\to\mu\mu\right)=\Gamma\left(B_s^0\left(t\right)\to\mu\mu\right)+\Gamma\left(\bar{B}_s^0\left(t\right)\to\mu\mu\right)$.  In the SM hypothesis $\taumumu$ coincides with the lifetime of the heavy $B_s^0$ eigenstate $\tau_{B_s^H}$.
The experimental average of the $B_s^0$ - $\bar{B}_s^0$ lifetimes and their difference~\cite{HFLAV:2022pwe} yields the prediction $\taumumu^\mathrm{SM}=\left( 1.624 \pm 0.009\right)\text{ ps}$, with new physics effects perturbing it at most by the difference between the heavy and light eigenstate lifetimes
($0.193$ ps \cite{HFLAV:2022pwe,PDG_2022}).

Previously, measurements of \taumumu\ (all largely consistent with the SM expectation) have been published by the CMS \cite{CMS-BPH-16-004, CMS-BPH-21-006} and LHCb \cite{LHCb_2017, LHCb_2022} collaborations, in conjunction with the latest results on the branching fractions $\mathrm{BR}\left(\B^0 \to\mu\mu\right)$
and $\mathrm{BR}\left(\B_s^0\to\mu\mu\right)$. Experimental results on these branching fractions have been published also by the ATLAS collaboration \cite{BPHY-2011-04,BPHY-2012-01,BPHY-2018-09}. A combination of the \taumumu\ measurements made by LHCb and CMS collaborations on their 2011-2016 datasets has also been published \cite{ATLAS-CONF-2020-049}.

The ATLAS experiment~\cite{PERF-2007-01} at the LHC is a multipurpose particle detector
with a forward--backward symmetric cylindrical geometry and a near \(4\pi\) coverage in
solid angle.\footnote{ATLAS uses a right-handed coordinate system
with its origin at the nominal interaction point. The $z$-axis is along
the beam pipe, the $x$-axis points to the centre of the LHC ring and
the $y$-axis points upward. Cylindrical coordinates $(r,\phi)$ are used
in the transverse plane, $r$ being the distance from the origin and
$\phi$ being the azimuthal angle around the beam pipe. The pseudorapidity
$\eta$ is defined as $\eta = - \ln[\tan (\theta/2)]$ where
$\theta$ is the polar angle.}
It consists of an inner tracking detector surrounded by a thin superconducting solenoid
providing a \qty{2}{\tesla} axial magnetic field, electromagnetic and hadron calorimeters, and a muon spectrometer.
The inner tracking detector covers the pseudorapidity range \(|\eta| < 2.5\).
It consists of silicon pixel, silicon microstrip, and transition radiation tracking detectors.
Lead/liquid-argon (LAr) sampling calorimeters provide electromagnetic (EM) energy measurements
with high granularity.
A steel/scintillator-tile hadron calorimeter covers the central pseudorapidity range (\(|\eta| < 1.7\)).
The endcap and forward regions are instrumented with LAr calorimeters
for both the EM and hadronic energy measurements up to \(|\eta| = 4.9\).
The muon spectrometer surrounds the calorimeters and is based on
three large superconducting air-core toroidal magnets with eight coils each.
The field integral of the toroids ranges between \num{2.0} and \qty{6.0}{\tesla\metre}
across most of the detector.
The muon spectrometer includes a system of precision tracking chambers and fast detectors for triggering.
A two-level trigger system is used to select events.
The first-level trigger is implemented in hardware and uses a subset of the detector information
to accept events at a rate below \qty{100}{\kHz}.
This is followed by a software-based trigger that
reduces the accepted event rate to \qty{1}{\kHz} on average
depending on the data-taking conditions.
An extensive software suite~\cite{ATL-SOFT-PUB-2021-001} is used in data simulation, in the reconstruction
and analysis of real and simulated data, in detector operations, and in the trigger and data acquisition
systems of the experiment.
 
\section{Dataset and event selection}
\label{sec:selection}
 
This analysis is based on the Run~2 data recorded in 2015
and 2016 from $pp$ collisions at the LHC at a center-of-mass energy of $\sqrt{s}=13$~\TeV.
Data used in the analysis were recorded during stable LHC beam periods.
Data quality requirements were imposed, notably on the
performance of the muon spectrometer, inner detector and calorimeter systems~\cite{DAPR-2018-01}.
The total integrated luminosity collected by ATLAS in this period is
$36.2$~fb$^{-1}$ with an uncertainty of 2.1\%.
These values are determined using a methodology similar to that detailed in Ref.~\cite{DAPR-2013-01},
based on calibration of the luminosity scale using $x$--$y$ beam-separation scans,
and use the LUCID-2 detector~\cite{LUCID_2_reference} for the baseline luminosity measurement.
 
The data and Monte Carlo (MC) samples used in this analysis are identical to those of the ATLAS collaboration's most recent publication on the \bmumu\ Branching Ratios (BR) \cite{BPHY-2018-09}, including
the same per-event MC weights employed to correct data-MC discrepancies.
 
Together with di-muon signal candidates, the \binjpsik\ signal
({\em reference channel}) is employed in the study of the analysis' systematic uncertainties. The data employed is collected with triggers subject to
time-dependent pre-scales~\cite{TRIG-2016-01}, affecting the signal and reference channel differently: accounting for
these effects, the total effective integrated luminosity employed amounts to $26.3$~fb$^{-1}$ for (\bsmumu\ ) and $15.1$~fb$^{-1}$ for (\binjpsik\ ).
 
$B$ meson candidates are reconstructed based on a decay vertex fitted to two or three tracks for the signal and the reference channels respectively, depending on the decay process to be reconstructed. The invariant mass of each $B$ candidate is calculated using muon
trajectories measured by combining the information from the inner detector and the muon spectrometer to improve upon the mass resolution
obtained from inner detector information only~\cite{PERF-2014-05}.
The coordinates of primary vertices are obtained from charged-particle tracks constrained to the luminous region of the colliding beams in the transverse plane, after excluding the tracks used in the signal candidates' reconstruction.
The matching of a $B$ candidate to a primary vertex is done by extrapolating the candidate trajectory
to the point of closest approach to the beam axis, and choosing the primary vertex with
the smallest separation along $z$.
Ref.~\cite{BPHY-2018-09} verifies in simulation that this method matches the correct vertex with a probability above $99\%$ for all relevant pile-up \footnote{The pile-up is defined as the average number of $pp$
collisions in the same bunch crossing} conditions.
The \bsmumu\ candidate mass is obtained from the measured di-muon four-momentum, while its proper decay time is calculated as  $\tilde{t}_{\mumu} =\frac{L_{xy} m_{B_s^0}^\mathrm{PDG}}{p_\mathrm{T}^{B_s^0}}$, where $L_{xy}$ is the decay length projected along the reconstructed
$B_s^0$ momentum in the transverse plane, $m_{B_s^0}^\mathrm{PDG}$ the world averaged mass of $B_s^0$ mesons from Ref.~\cite{PDG_2022} and $p_\mathrm{T}^{B_s^0}$ the magnitude of the candidate's reconstructed transverse momentum.

The same analysis selections as in Ref.~\cite{BPHY-2018-09} are employed, including the same definition for the invariant mass signal-dominated ($\left[5166 - 5526\right]$ ~MeV) and background-dominated ($\left[4766-5166 \right]$ ~MeV and $\left[5526 - 5966\right]$ ~MeV) sideband regions. Invariant mass fits are performed in the $\left[4766 - 5966\right]$ ~MeV range.
As part of the event selection, the Boosted Decision Tree (BDT) from Ref.~\cite{BPHY-2018-09} is employed to discriminate signal from the very large background: this BDT relies on 15 physical input variables related either to isolation properties of the $B$
candidates and final state particles, or to topological and kinematic properties of the \bsmumu\, decay (see Ref.~\cite{BPHY-2018-09} for details).
 
The final event selection is simplified from multiple BDT output categories of Ref.~\cite{BPHY-2018-09} to a single one. The requirement is chosen optimising
the signal significance $\frac{S}{\sqrt{S+B}}$ in the
$\left[5166 - 5526\right]$ ~MeV candidate invariant mass range. The optimisation procedure is based on MC simulated events normalised to the SM expectation for the signal yield ($S$), and to background in a looser-BDT control region for the background yield ($B$).
The background control region coincides with the  {\em bin 0} BDT requirement (>0.1439) employed in the BR analysis~\cite{BPHY-2018-09}.
This requirement selects a background-dominated sample immediately next to the signal-sensitive BDT region. The control region is employed exclusively in this optimisation procedure.
 
The optimal BDT range is found to be $\left[0.3650 - 1\right]$, corresponding to 49 signal and 27 background events expected - under the SM hypothesis - in the $\left[5166 - 5526\right]$ ~MeV candidates' invariant mass range.
 
\section{The effective lifetime measurement}
 
\begin{figure}[htb]
\centering
\includegraphics[width=0.95\textwidth]{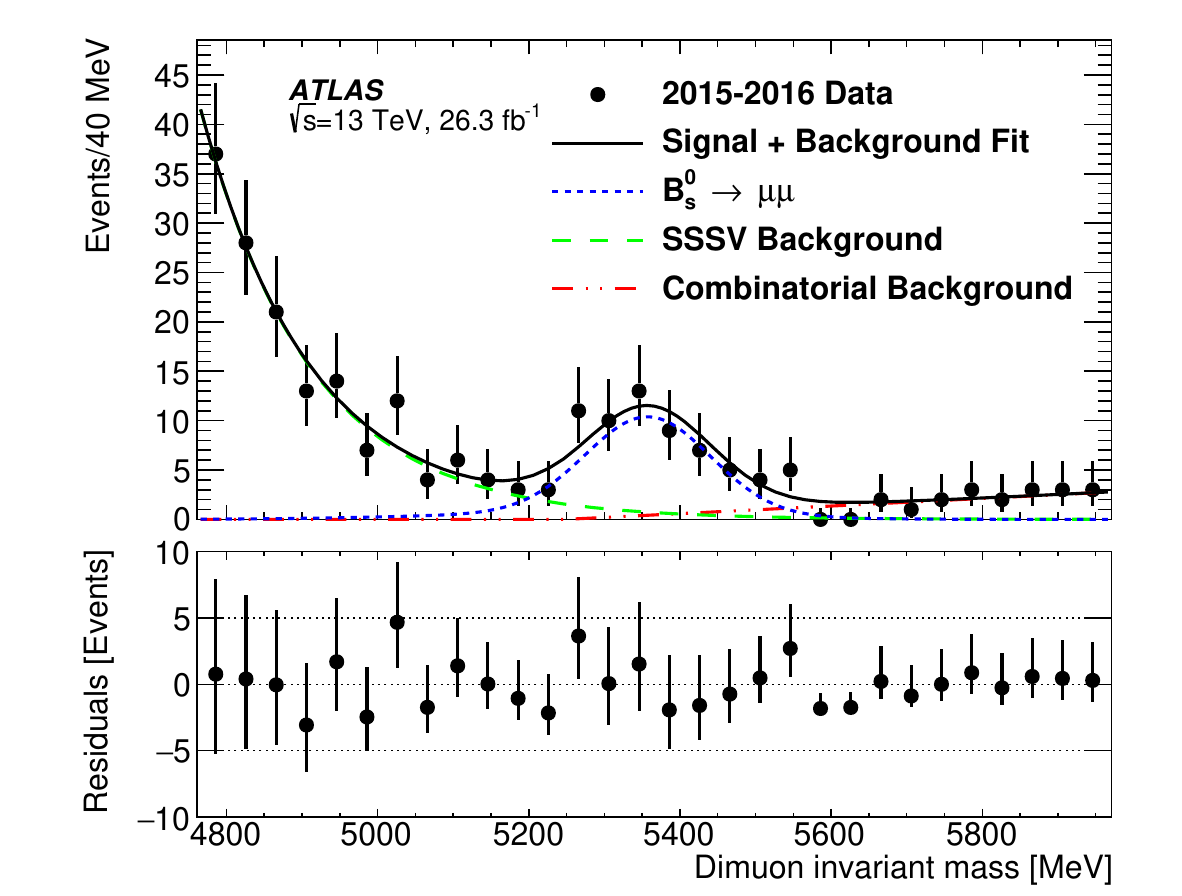}
\caption{Invariant mass distribution of di-muon candidates passing the optimised BDT selection in the 2015-2016 dataset.
The linear background component (red) is consistent in shape with simulations of di-muons originating from different $b$-quarks in $pp\to b\bar{b}$ processes.
The exponentially falling background component (green) is instead consistent in shape with {\it Same Side Same Vertex} (SSSV) decays, i.e. muons corresponding
to particles produced from the decay of a single $b$ or $\bar{b}$ quark. The $B_s^0$ signal component is shown in blue.\\
The bottom panel reports the per-bin residuals, calculated as the difference between the top panel's data points and the black fit line.}
\label{fig:dataInvariantMassPlot}
\end{figure}

After the event selection outlined in Section~\ref{sec:selection}, the invariant mass distribution of actual data candidates is shown in Figure~\ref{fig:dataInvariantMassPlot}. The superimposed five parameters invariant mass fit is the result of an un-binned extended maximum likelihood fit to candidates in the $\left[ 4766 -5966\right]$ ~MeV mass region.
The fit includes three PDF models, identical to what was used in the BR analysis~\cite{BPHY-2018-09} (which motivates the choice through MC studies). A
double-Gaussian is employed to model the \bsmumu\, signal. The background model includes an exponentially decaying component (to model {\em same-side same-vertex} (SSSV) $B$ meson decays, i.e. muon pairs originating from a single $B$ meson decay) and a linear contribution (to model the {\em combinatorial background}, i.e. random pairing of muons from $b\bar{b}$ decays).
The latter component has a free slope and is imposed to be zero whenever the linear model would become negative.
All yields and background shape parameters are unconstrained in the fit. Consistently with Ref.\cite{BPHY-2018-09}, the relative fraction, mean and width of
the signal Gaussian distributions are constrained to the fully simulated MC signal, within the detector mass-scale ($\pm5$~MeV) and
mass-resolution ($\pm5\,\%$) uncertainties.
Additional resonant and non-resonant contributions are neglected in the fit model. The corresponding systematic uncertainties are evaluated by comparing the
measurement shift using simulations that include these additional contributions, as explained in Section~\ref{sec:syst}.
The fit yields $58\pm13$ (stat. only) \bsmumu\ signal events in the range $\left[4766-5966 \right]$ ~MeV.
 
\begin{figure}[htb]
\centering
\includegraphics[width=0.95\textwidth]{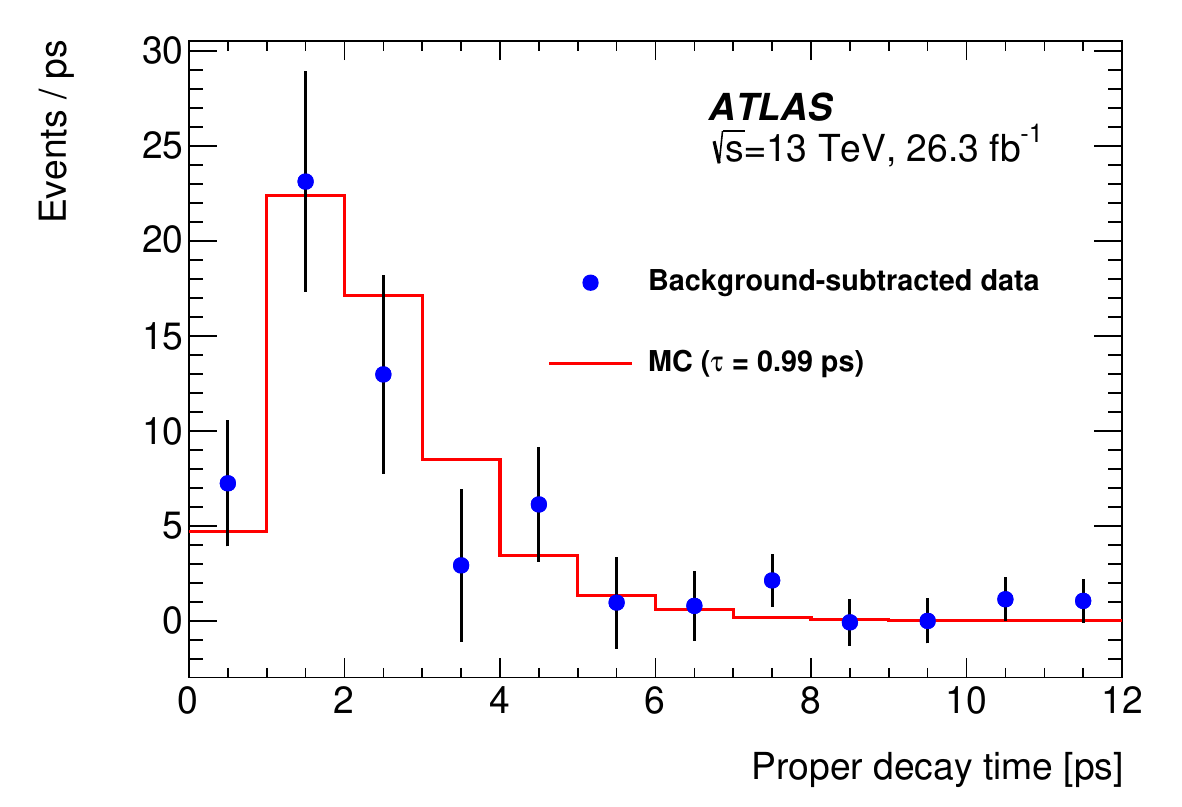}
\caption{Signal proper decay time distribution extracted with the \textit{sPlot} background subtraction procedure applied to the invariant mass fit illustrated in Figure \ref{fig:dataInvariantMassPlot}. The superimposed signal MC template is the result of the
lifetime fit procedure discussed in the text. The uncertainties on the data points are calculated as Poisson fluctuations on the MC yield prediction (continuous red histogram) in the corresponding bin.}
\label{fig:dataSignalPlot}
\end{figure}

The proper decay time distribution in data is background-subtracted employing per-event weights calculated according to the {\it sPlot} technique \cite{sPLOT}: signal and backgrounds weights are calculated from the result of the mass fit, yielding the
background-subtracted distribution shown in Figure \ref{fig:dataSignalPlot}. The lifetime measurement is obtained by minimising the binned $\chi^2$ between the data histogram and lifetime-dependent pure signal MC templates extracted from MC simulated samples, as illustrated below in Figure~\ref{fig:Chi2Scan}. The $\chi^2$ calculation accounts for the statistical uncertainty on the weight-corrected MC as well as the Poissonian uncertainty in each data bin as expected from the predicted MC content for that bin.
The MC templates, corresponding to different lifetimes, are generated as a function of the fit parameter \taumumuobs\ by re-weighting each signal MC event by the signal true proper decay time distribution.
The evaluation of the $\tau_{\mu\mu}$ statistical sensitivity and of most systematic uncertainties relies on MC pseudo-experiments. These are based on generating
the events in the phase space of di-muon mass and proper decay time, using analytical models for the signal and background components.
The mass parameterisation of each of these components is consistent with the invariant mass fit, while
the proper decay time distributions are empirically modelled with an exponential function (weighted by an error function in the case of the signal component) to take into account acceptance and efficiency effects. Finally, the distributions are convoluted with a Gaussian Probability Density Function (PDFs), representing the time resolution.
The parameters of these models are fixed according to the signal and background components in MC simulated samples. The resulting shapes are verified to be consistent with the signal and background distributions obtained from data.
 
When extracting the \bsmumu\ lifetime, the bin width of the proper decay time histogram is chosen for simplicity to be constant. The analysis procedure is applied to MC pseudo-experiments for different bin widths and fit ranges, choosing the optimal configuration to reach the
best statistical uncertainty on \taumumuobs\ .
A closure test is performed on MC pseudo-experiments generated at $\taumumu^\mathrm{True}=\taumumu^\mathrm{SM}$, yielding a bias on \taumumuobs\ of $(82 \pm 4) \text{ fs}$. This bias is verified to arise from the low-statistics regime of the fit.
\taumumuobs\  is therefore redefined as $\taumumuobs-82$~fs and taken as the central value for the
measurement, with the uncertainty estimated from a MC pseudo-experiments based Neyman construction \cite{neyman}. As $\taumumu^\mathrm{True}$ is varied in the range $\left[\tau_{B_s^L},\tau_{B_s^H}\right]$ a 15~fs bias decrease is observed.
This value is included as systematic uncertainty due to the fit bias lifetime dependency (see Section~~\ref{sec:syst}).

\begin{figure}[htb]
\centering
\includegraphics[width=0.95\textwidth]{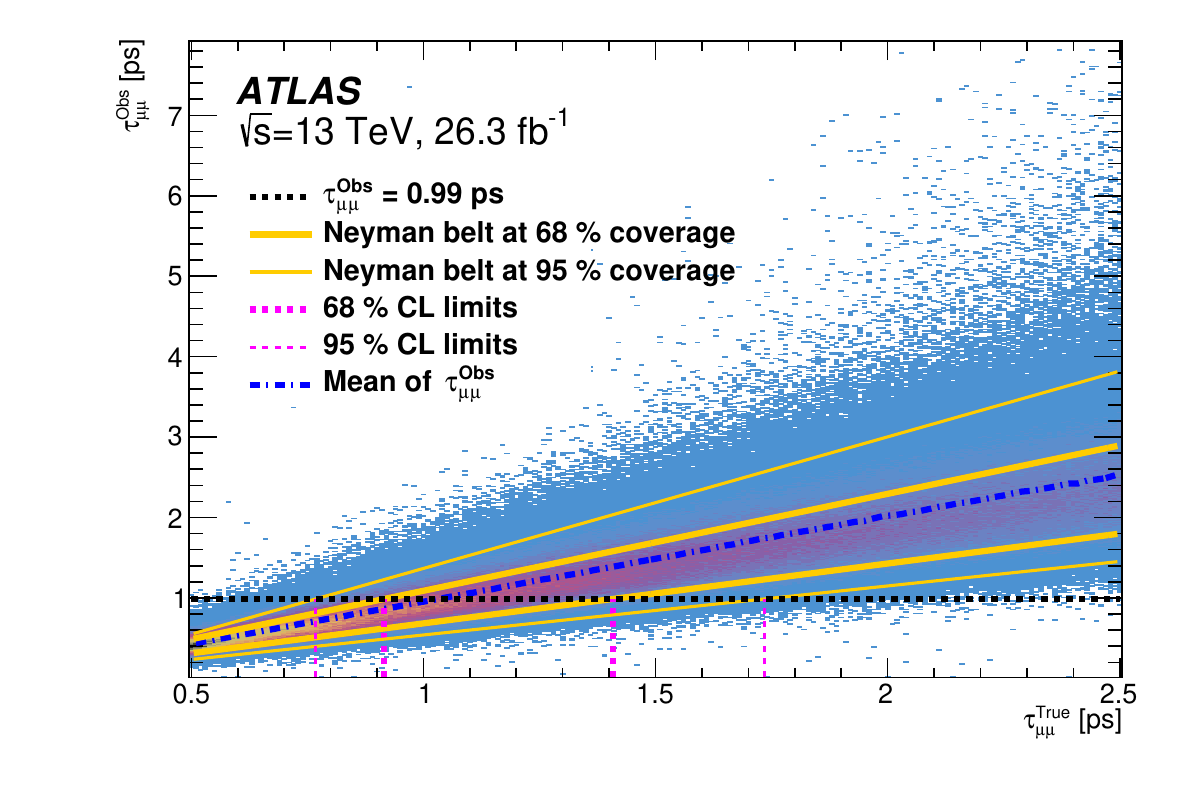}

\caption{$68\%$ and $95\%$ CL bands obtained with a Neyman construction based on MC pseudo-experiments for the signal and background components. The yellow
lines interpolate the band boundaries in order to smooth the effects of limited number of MC pseudo-experiments used.
The dashed-dotted blue line corresponds to the average expected $\tau_{\mu\mu}^\mathrm{Obs}$ value at a given $\tau_{\mu\mu}^\mathrm{True}$ value. The
horizontal dashed black line corresponds to the experimentally observed value of $\tau_{\mu\mu}^\mathrm{Obs}=0.99 \text{ ps}$,
yielding a  68$\%$ CL band of $[0.92,1.41] \text{ ps}$ (thick vertical dashed purple lines) and a 95$\%$ CL band of $[0.77,1.73] \text{ ps}$ (thin vertical dashed purple lines).
The same construction at the $\tau_{\mu\mu}^\mathrm{Obs}$ corresponding to $\tau_{\mu\mu}^\mathrm{True}=\tau_{\mu\mu}^\mathrm{SM}$ yields $[1.44,2.26] \text{ ps}$ as  68$\%$ CL band.}
\label{fig:toyNeymanBands}
\end{figure}

Figure~\ref{fig:Chi2Scan} reports the $\chi^2$ scan as a function of \taumumuobs. 
The $\Delta\chi^2=1$ interval from this curve would not result in a reliable estimation of the \taumumuobs\ statistical uncertainty with the non-Gaussian estimator used for the measurement. This uncertainty is instead derived from the Neyman CL band construction
illustrated in Figure~\ref{fig:toyNeymanBands}. The $\chi^2$ minimum and the Neyman belt construction yield
$\taumumuobs=0.99 ^{+0.42}_{-0.07} \text{ (stat. only)}\,\mathrm{ps}$. The imbalance between positive and negative statistical uncertainties is already suggested by the asymmetry in the $\chi^2$ scan of Figure~\ref{fig:Chi2Scan}.
The effect is further emphasised by the subtraction of the closure test bias.

\begin{figure}[h]
\centering
\includegraphics[width=0.8\textwidth]{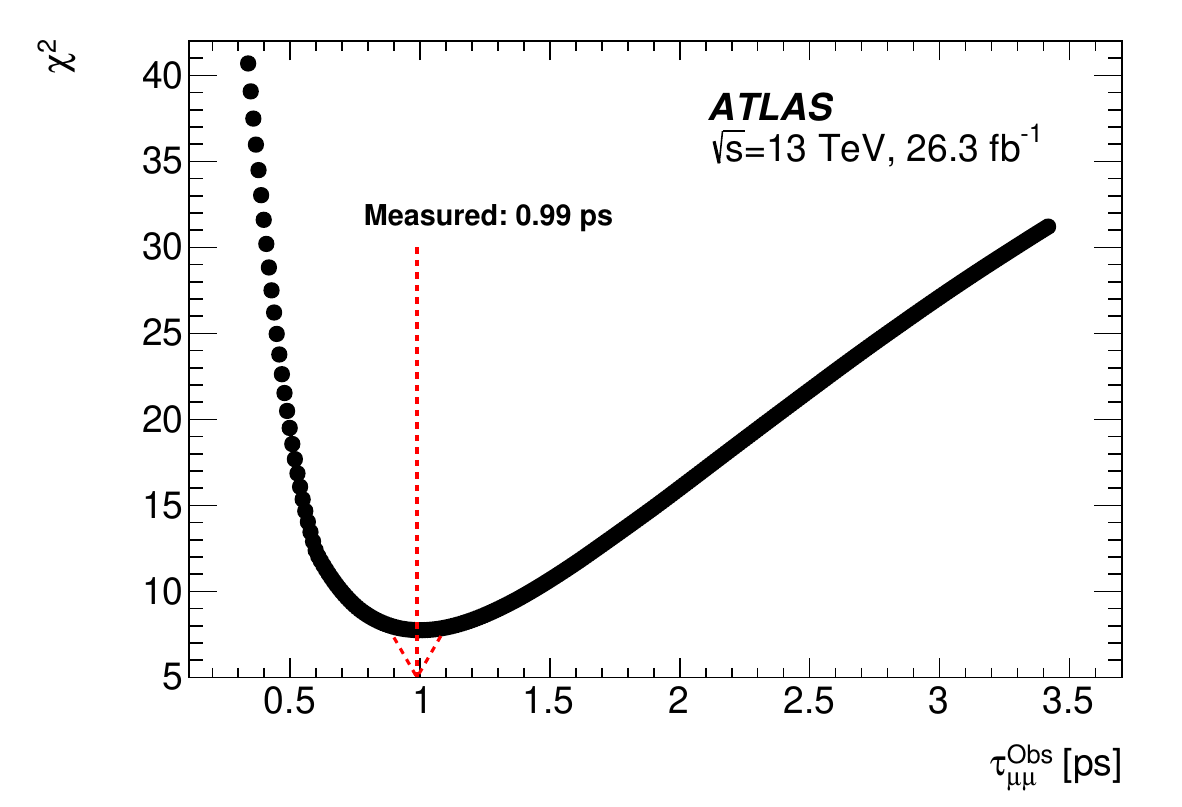}
\caption{$\chi^2$ scan vs MC lifetime. The minimum of the scan ($\chi^2/\mathrm{NDOF}=7.7/11$), located at $\tau_{\mu\mu}^\mathrm{Obs}=0.99$ ps, is used to determine the central value of the lifetime measured and is indicated by the red vertical dashed
arrow.}
\label{fig:Chi2Scan}
\end{figure}

\FloatBarrier

\section{Systematic uncertainties}
\label{sec:syst}
 
Systematic uncertainties on \taumumu\ arise from fit-procedure assumptions, data-MC discrepancies and neglected backgrounds. These effects are discussed in detail below and estimated, unless otherwise specified, with the MC pseudo-experiments described above.
 
First, the fit procedure is based on a number of assumptions.
The analytical models describing the {\em SSSV} and {\em combinatorial} backgrounds are replaced respectively with a Gaussian tail and an exponential function, yielding average shifts of
22 fs and 14 fs. A shift of 60 fs is observed when the number of {\em SSSV} background events is varied by $\pm100\%$ in the simulation to account for normalisation assumptions.
This variation is used without any further refinement as it is quite small compared to the expected statistical uncertainty although it is conservative with respect to the SSSV yield uncertainty ($ 152 \pm 13$ events from the fit on data).
 
The {\it sPlot} re-weighting effectively subtracts the combinatorial background relying on an admixture of data events above and below the signal peak invariant mass region. A potential correlation between the background candidates' proper decay time and invariant mass is tested by repeating the fit and {\it sPlot} extraction of \taumumu\ on the same data, excluding in turn the upper or the lower sidebands. The largest shift observed for these two options is $56$ fs and is taken as systematic uncertainty.
 
The nominal invariant mass fit does not take into account other $b$-hadron decays whose presence is considered as a source of systematic uncertainty. Each of these contributions is individually merged (in proportion to its expected SM yield after the analysis selection) with
the normal MC pseudo-experiments, and the average difference in measured lifetime before and after this inclusion is measured for the semi-leptonic $b$-meson decays ($2 \text{ fs}$), the two-body hadronic $b$-meson decays ($3 \text{ fs}$),
the inclusive $B_c^{\pm}$ decays ($10 \text{ fs}$) and the $B^0 \to \mu\mu$ decays ($16 \text{ fs}$).
 
The dominant data-MC systematic effect arises from the difference in vertex resolution between data and MC. This resolution tends to be underestimated in MC but is also distributed differently for the signal and reference channels.
The effect is therefore estimated by measuring the \BpmKpmJpsi\ lifetime on data and comparing the result against the world-average experimental value \cite{HFLAV:2022pwe}. The measurement is performed applying to the reference channel the same fit procedure
employed to extract \taumumu\  . The average difference between the result obtained on \BpmKpmJpsi\ data and MC pseudo-experiments is then measured in bins of proper decay
length resolution ($\sigma_{L_{xy}}$). A bin-by-bin weighted average based on the proper decay length resolution distribution of the simulated \bsmumu\ signal is performed to take
into account differences between signal and reference channels. The final shift is found to be $134 \text{ fs}$. Aside from topological differences between the reference and signal channels, kinematical differences can also skew the measurement. The
data/MC ratio for the \BpmKpmJpsi\ signal di-muon pseudo-rapidity separation is applied to the \bsmumu\ signal, yielding an additional shift of $6 \text{ fs}$.  Uncertainties in the kinematic and reconstruction corrections (detailed in Ref. \cite{BPHY-2018-09}) applied
to MC candidates are accounted for by repeating the measurement on MC pseudo-experiments with and without these corrections applied, observing a combined shift of $65 \text{ fs}$.
 
In this analysis the MC proper decay time fit templates are derived with the assumption that only the heavy $B^0_s$ mass eigenstate contributes to the decay, as predicted by the SM. The systematic effect due to this assumption is already included in the  $15 \text{ fs}$ systematic uncertainty ascribed to the fit bias lifetime dependency.

All the systematic effects discussed above are conservatively symmetrised and then combined in quadrature into an overall systematic uncertainty of $0.17 \text{ ps}$, yielding the ATLAS measurement of
$\taumumuobs=0.99 ^{+0.42}_{-0.07} \text{ (stat.)} \pm 0.17 \text{ (syst.)}\,\mathrm{ps}$. The impact on \taumumu\ of the different sources of systematic uncertainties described above is reported in Table~\ref{tab:systs}.
 
\begin{table}[h]
\centering
\begin{tabular}{c|c}
\textbf{Uncertainty source} & \textbf{$\Delta \taumumuobs\ $} [fs] \\\hline
Data - MC discrepancies & 134 \\
SSSV lifetime model & 60\\
Combinatorial lifetime model & 56\\
$B$ kinematic reweighting & 55\\
$B$ isolation reweighting & 32\\
SSSV mass model & 22\\
$B_d$ background & 16\\
Fit bias lifetime dependency and $B^0_s$ eigenstates admixture & 15\\   
Combinatorial mass model & 14\\
Pileup reweighting & 13 \\
$B_c$ background & 10\\
Muon $\Delta_\eta$ correction & 6\\  
$B\to hh'$ background & 3\\
Muon reconstruction SF reweighting & 2 \\  
Semileptonic background & 2\\
Trigger reweighting & 1\\\hline
\textbf{Total} & 174\\\hline   
\end{tabular}
\caption{\label{tab:systs} Summary of the systematic uncertainty contributions
affecting the \bsmumu\ lifetime measurement. The last line
represents the sum in quadrature of all systematic effects.}
\end{table}

\section{Conclusions}
This paper presents the first ATLAS measurement of the \bsmumu\ effective lifetime, based on a fraction of the experiment's Run 2 dataset corresponding to 26.3 fb$^{-1}$ of 13 TeV LHC proton-proton collisions.
The result obtained is $\taumumuobs=0.99 ^{+0.42}_{-0.07} \text{ (stat.)} \pm 0.17 \text{ (syst.)}\,\mathrm{ps}$. It is consistent with the SM prediction $\taumumu^\mathrm{SM}=\left( 1.624 \pm 0.009\right)\text{ ps}$~\cite{HFLAV:2022pwe} as well as with the other available experimental results.

\clearpage


\section*{Acknowledgements}


We thank CERN for the very successful operation of the LHC, as well as the
support staff from our institutions without whom ATLAS could not be
operated efficiently.
 
We acknowledge the support of
ANPCyT, Argentina;
YerPhI, Armenia;
ARC, Australia;
BMWFW and FWF, Austria;
ANAS, Azerbaijan;
CNPq and FAPESP, Brazil;
NSERC, NRC and CFI, Canada;
CERN;
ANID, Chile;
CAS, MOST and NSFC, China;
Minciencias, Colombia;
MEYS CR, Czech Republic;
DNRF and DNSRC, Denmark;
IN2P3-CNRS and CEA-DRF/IRFU, France;
SRNSFG, Georgia;
BMBF, HGF and MPG, Germany;
GSRI, Greece;
RGC and Hong Kong SAR, China;
ISF and Benoziyo Center, Israel;
INFN, Italy;
MEXT and JSPS, Japan;
CNRST, Morocco;
NWO, Netherlands;
RCN, Norway;
MEiN, Poland;
FCT, Portugal;
MNE/IFA, Romania;
MESTD, Serbia;
MSSR, Slovakia;
ARRS and MIZ\v{S}, Slovenia;
DSI/NRF, South Africa;
MICINN, Spain;
SRC and Wallenberg Foundation, Sweden;
SERI, SNSF and Cantons of Bern and Geneva, Switzerland;
MOST, Taiwan;
TENMAK, T\"urkiye;
STFC, United Kingdom;
DOE and NSF, United States of America.
In addition, individual groups and members have received support from
BCKDF, CANARIE, Compute Canada and CRC, Canada;
PRIMUS 21/SCI/017 and UNCE SCI/013, Czech Republic;
COST, ERC, ERDF, Horizon 2020 and Marie Sk{\l}odowska-Curie Actions, European Union;
Investissements d'Avenir Labex, Investissements d'Avenir Idex and ANR, France;
DFG and AvH Foundation, Germany;
Herakleitos, Thales and Aristeia programmes co-financed by EU-ESF and the Greek NSRF, Greece;
BSF-NSF and MINERVA, Israel;
Norwegian Financial Mechanism 2014-2021, Norway;
NCN and NAWA, Poland;
La Caixa Banking Foundation, CERCA Programme Generalitat de Catalunya and PROMETEO and GenT Programmes Generalitat Valenciana, Spain;
G\"{o}ran Gustafssons Stiftelse, Sweden;
The Royal Society and Leverhulme Trust, United Kingdom.
 
The crucial computing support from all WLCG partners is acknowledged gratefully, in particular from CERN, the ATLAS Tier-1 facilities at TRIUMF (Canada), NDGF (Denmark, Norway, Sweden), CC-IN2P3 (France), KIT/GridKA (Germany), INFN-CNAF (Italy), NL-T1 (Netherlands), PIC (Spain), ASGC (Taiwan), RAL (UK) and BNL (USA), the Tier-2 facilities worldwide and large non-WLCG resource providers. Major contributors of computing resources are listed in Ref.~\cite{ATL-SOFT-PUB-2021-003}.


\clearpage

\printbibliography
 
\clearpage
 
\begin{flushleft}
\hypersetup{urlcolor=black}
{\Large The ATLAS Collaboration}

\bigskip

\AtlasOrcid[0000-0002-6665-4934]{G.~Aad}$^\textrm{\scriptsize 102}$,
\AtlasOrcid[0000-0002-5888-2734]{B.~Abbott}$^\textrm{\scriptsize 120}$,
\AtlasOrcid[0000-0002-1002-1652]{K.~Abeling}$^\textrm{\scriptsize 55}$,
\AtlasOrcid[0000-0001-5763-2760]{N.J.~Abicht}$^\textrm{\scriptsize 49}$,
\AtlasOrcid[0000-0002-8496-9294]{S.H.~Abidi}$^\textrm{\scriptsize 29}$,
\AtlasOrcid[0000-0002-9987-2292]{A.~Aboulhorma}$^\textrm{\scriptsize 35e}$,
\AtlasOrcid[0000-0001-5329-6640]{H.~Abramowicz}$^\textrm{\scriptsize 151}$,
\AtlasOrcid[0000-0002-1599-2896]{H.~Abreu}$^\textrm{\scriptsize 150}$,
\AtlasOrcid[0000-0003-0403-3697]{Y.~Abulaiti}$^\textrm{\scriptsize 117}$,
\AtlasOrcid[0000-0002-8588-9157]{B.S.~Acharya}$^\textrm{\scriptsize 69a,69b,q}$
% ...
\bigskip
\\

% ...
$^{29}$Physics Department, Brookhaven National Laboratory, Upton NY; United States of America.\\
% ...
$^{35}$$^{(a)}$Facult\'e des Sciences Ain Chock, R\'eseau Universitaire de Physique des Hautes Energies - Universit\'e Hassan II, Casablanca;$^{(b)}$Facult\'{e} des Sciences, Universit\'{e} Ibn-Tofail, K\'{e}nitra;$^{(c)}$Facult\'e des Sciences Semlalia, Universit\'e Cadi Ayyad, LPHEA-Marrakech;$^{(d)}$LPMR, Facult\'e des Sciences, Universit\'e Mohamed Premier, Oujda;$^{(e)}$Facult\'e des sciences, Universit\'e Mohammed V, Rabat;$^{(f)}$Institute of Applied Physics, Mohammed VI Polytechnic University, Ben Guerir; Morocco.\\
% ...
$^{49}$Fakult\"{a}t Physik , Technische Universit{\"a}t Dortmund, Dortmund; Germany.\\
% ...
$^{55}$II. Physikalisches Institut, Georg-August-Universit\"{a}t G\"ottingen, G\"ottingen; Germany.\\
% ...
$^{69}$$^{(a)}$INFN Gruppo Collegato di Udine, Sezione di Trieste, Udine;$^{(b)}$ICTP, Trieste;$^{(c)}$Dipartimento Politecnico di Ingegneria e Architettura, Universit\`a di Udine, Udine; Italy.\\
% ...
$^{102}$CPPM, Aix-Marseille Universit\'e, CNRS/IN2P3, Marseille; France.\\
% ...
$^{117}$Department of Physics, New York University, New York NY; United States of America.\\
% ...
$^{120}$Homer L. Dodge Department of Physics and Astronomy, University of Oklahoma, Norman OK; United States of America.\\
% ...
$^{150}$Department of Physics, Technion, Israel Institute of Technology, Haifa; Israel.\\
$^{151}$Raymond and Beverly Sackler School of Physics and Astronomy, Tel Aviv University, Tel Aviv; Israel.\\
% ...
$^{a}$ Also Affiliated with an institute covered by a cooperation agreement with CERN.\\
$^{b}$ Also at An-Najah National University, Nablus; Palestine.\\
$^{c}$ Also at APC, Universit\'e Paris Cit\'e, CNRS/IN2P3, Paris; France.\\
$^{d}$ Also at Borough of Manhattan Community College, City University of New York, New York NY; United States of America.\\
$^{e}$ Also at Center for High Energy Physics, Peking University; China.\\
$^{f}$ Also at Center for Interdisciplinary Research and Innovation (CIRI-AUTH), Thessaloniki; Greece.\\
% ...
$^{q}$ Also at Department of Physics, King's College London, London; United Kingdom.\\
% ...

\end{flushleft}


\end{document}